\documentclass[pra,twocolumn,preprintnumbers,amsmath,amssymb,nofootinbib,showpacs]{revtex4-1}
\usepackage{graphicx}
\usepackage{amssymb}
\usepackage{mathrsfs}
\usepackage{color}

\newcommand{\beq}{\begin{equation}}
\newcommand{\eeq}{\end{equation}}

\begin{document}

\title{Galilean invariance at quantum Hall edge}

\author{Sergej Moroz$^{1,2}$, Carlos Hoyos$^3$ and Leo Radzihovsky$^{1,2}$}
\affiliation{$^1$Department of Physics, University of Colorado, Boulder, Colorado 80309, USA \\
                $^2$Center for Theory of Quantum Matter, University of Colorado, Boulder, Colorado 80309, USA \\
                $^3$Department of Physics, Universidad de Oviedo, Avda. Calvo Sotelo 18, 33007, Oviedo, Spain}


%
%

\begin{abstract}
We construct the theory of a chiral Luttinger liquid that lives on the boundary of a Galilean invariant quantum Hall fluid. In contrast to previous studies, Galilean invariance of the total (bulk plus edge) theory is guaranteed. We consider electromagnetic response at the edge and calculate momentum- and frequency-dependent electric conductivity and argue that its experimental measurement can provide a new means to determine the ``shift'' and bulk Hall viscosity. 
\end{abstract}

\pacs{73.43.-f}

\maketitle

\section{Introduction} \label{intro}
Despite a long history and deep  understanding of many microscopic and universal features of quantum Hall states \cite{Ezawa2000,Fradkin2013}, quantum Hall physics continues to present experimental surprises and new theoretical challenges. Recent theoretical  developments regarding the ``shift'' \cite{wen1992},  the Hall viscosity \cite{2009PhRvB..79d5308R,2011PhRvB..84h5316R}, its relation to electromagnetic response \cite{Hoyos2012,Bradlyn2012,Son2013} and generic geometric response \cite{Bradlyn2014,Gromov2014c, Can2014,Gromov2015} prove that there is still a lot to be learnt about quantum Hall fluids. Motivated by these advances and the intimate connection between the quantum Hall fluid's bulk and edge properties \cite{Halperin1982,KaneFisher,Fradkin2013}, we consider here a finite droplet of a Galilean invariant quantum Hall fluid
and  reexamine the physics that takes place at its edge.

\begin{figure}[ht]
\begin{center}
\includegraphics[height=0.25\textwidth]{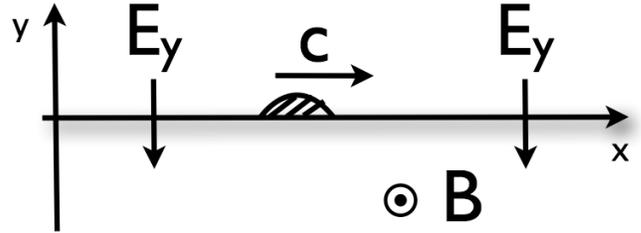}
\caption{Chiral excitation (hatched bump) that propagates with the drift velocity $c_x=-E_y/B$ along the edge ($y=0$) of a quantum Hall droplet. }\label{fig1}
\end{center}
\end{figure}

We will study a quantum Hall fluid that is incompressible in the bulk. At the edge, however, it supports gapless excitations which are long wavelength deformations in the shape of the droplet that propagate along the boundary with a drift velocity. The standard theory of edge states, put forward in the seminal papers of Wen and Stone \cite{Wen1992a,Stone1991}, is by now well-established and understood \cite{KaneFisher,Fradkin2013}. It is a theory of a \emph{chiral Luttinger liquid}, where excitations propagate along the edge only in one direction that is determined by the direction of the magnetic field (see Fig. \ref{fig1}). Tunneling experiments confirmed the chiral Luttinger liquid nature of the edge states in quantum Hall fluids \cite{Chang2003}. 

A remarkable property of the chiral Luttinger liquid is that it has an \emph{electromagnetic gauge anomaly}. Namely, taken on its own in the absence of the bulk, the edge theory does not make good sense because it is not gauge invariant under electromagnetic $U(1)$ transformations. 
As a result, in the presence of an electric field $E_x$ pointing along the boundary the electromagnetic current is \emph{not conserved}. The nonconservation of the edge current is compensated by the inflow of the Hall current from the gapped bulk, derived from the Chern-Simons term, via the celebrated Callan-Harvey mechanism \cite{Callan1985}. In this way the gauge invariance of the total system (bulk plus edge) is guaranteed.

Nevertheless, it turns out however that the standard edge theory has some deficiencies. In particular, we will show that the edge theory \cite{Wen1992a,Stone1991} is not invariant under Galilean boosts along a flat boundary in the presence of an external electromagnetic field. Since the bulk effective theory, encoded in the Chern-Simons action, is Galilean invariant, this implies the absence of Galilean invariance in the total (bulk plus boundary) theory. We conclude thus that for a clean Galilean invariant quantum Hall fluid the edge theory of Wen and Stone is incomplete.

To cure this deficiency it is helpful to impose a broader set of general coordinate invariance of which the Galilean invariance of interest to us is a subset.
General coordinate invariance is a central principle in Einstein's formulation of general theory of relativity. A decade ago in a seminal paper Son and Wingate extended this principle to nonrelativistic physics \cite{Son2006}. Nonrelativistic general coordinate invariance led to exciting and experimentally verifiable predictions in condensed matter and cold atom physics \cite{Son2006,Son:2007,Hoyos2012,Hoyos2013}. In technical terms, we will impose invariance of the edge theory under spatial diffeomorphism transformations, i.e. coordinate reparametrizations. In principle this allows to investigate the edge chiral Luttinger liquid living on curved boundaries of quantum Hall fluids placed on an arbitrary two-dimensional manifold, but in this paper we will concentrate our attention on the flat space physics. In particular, general coordinate invariance together with electromagnetic gauge invariance will guarantee the Galilean invariance of the total (bulk plus boundary) theory.

In addition to the formulation of a Galilean invariant theory of the chiral quantum Hall edge, our key result is its application to compute the linear response to an external electromagnetic perturbation. Namely, we find that the finite-frequency ($\omega$) and momentum ($p_x$) electric conductivity is given by 
\beq
\sigma(\omega, p_x) = \frac{\nu}{2 \pi} \Big( 1 +\frac{\mathcal{S}}{4} \frac{p^2_x}{B} \Big) \frac{-i c }{\omega- c p_x+i0^+} -im \epsilon''(B)\frac{p_x}{B},
\eeq
where $B$ is the magnetic field, $c$ the speed of the edge modes and $m$ the mass of the particles. The first term of the conductivity is universal because it is determined only by the filling factor $\nu$ and the ``shift'' $\mathcal{S}$. On the other hand, the second term is not universal since it is sensitive to the equation of state, i.e., energy density $\epsilon$ as a function of the magnetic field. Due to the close relation between the ``shift'' and the bulk Hall viscosity \cite{2009PhRvB..79d5308R,2011PhRvB..84h5316R}, a spectroscopic measurement of the momentum dependence of the conductivity at the edge provides an alternative way to experimentally determine the bulk Hall viscosity.
Our results are thus consistent with the bulk calculation of the electromagnetic response of quantum Hall fluids \cite{Hoyos2012,Bradlyn2012,Son2013}. 

In this paper we investigate an edge of integer quantum Hall and Laughlin (filling factor $\nu=1/k$ with $k$ an odd integer) fractional quanutm Hall fluids of spinless (spin polarized) fermions. It is known that in this case there is a \emph{single} chiral mode propagating along the boundary. Here we will not consider an edge of the hierarchical states of Haldane and Halperin \cite{Haldane1983,Halperin1984}, where \emph{multiple} edge modes are expected and the effects of disorder are important \cite{KaneFisher}. 

Predictions made in this paper should be of relevance for quantum Hall states to be realized in clean (no disorder) heterojunctions and rotating quantum degenerate fermionic atoms.

\section{Chiral Luttinger liquid} \label{CLL}
A convenient description of a quantum Hall edge is bosonization of a chiral Luttinger liquid.
In the integer quantum Hall fluid the edge excitations are free chiral fermions and the effective action can be readily written down. In the fractional quantum Hall case the fermionic formulation can not be studied by perturbative methods. In both cases however the theory can be bosonized.

Specifically, from the fermionic edge density $\rho$ we introduce first a chiral bosonic field $\theta$ by
\beq \label{rhoeq}
\rho=\frac{1}{2\pi}\partial_x \theta,
\eeq
where the coupling to the electromagnetic field $A_\mu$ has been neglected for the moment. Using this bosonic field, the edge theory can be expressed as \cite{Wen1992a,Stone1991}
\begin{equation}
S_\theta=\frac{1}{4\pi\nu}\int d^2 x\, \left(\partial_t \theta +c\partial_x \theta\right)\partial_x\theta,
\end{equation}
where $d^2x\equiv dt dx$ and $c$ is the velocity of the edge excitation. For a smooth edge, it is determined, in the simplest approximation, by the potential that confines the droplet and is given by the drift velocity of fermions located at the edge
\begin{equation} \label{drift}
c=-\frac{E_y}{B},
\end{equation}
where the electric field is $E_i\equiv \partial_t A_i-\partial_i A_t$ and the magnetic field is $B \equiv \epsilon^{ij}\partial_i A_j$ with $\epsilon^{xy}=+1$.
In the bosonized formulation the effect of the local fermion interaction term $v_{int} \rho^2\sim v_{int} (\partial_x \theta)^2$ can be simply absorbed into the definition of the velocity $c$ \cite{Fradkin2013,KaneFisher}.
 
We consider now how the chiral boson couples to the external electromagnetic field $A_{\mu}$ neglected above. To this end, 
we derive the edge action from the bulk theory following the argument of Wen \cite{Wen1992a} in the presence of $A_{\mu}$.
First, we start with the effective action of a quantum Hall fluid which is the Chern-Simons action for a dynamical \emph{statistical gauge field} $a_\mu$ coupled to the external electromagnetic field  $A_\mu$
\begin{equation} \label{QH}
S_{QH}=-\frac{1}{4\pi\nu}\int d^3 x\, \epsilon^{\mu\nu\lambda}a_\mu\partial_\nu a_\lambda-\frac{1}{2\pi}\int d^3 x\, \epsilon^{\mu\nu\lambda}a_\mu\partial_\nu A_\lambda.
\end{equation}
The action was written in such a way that it is explicitly invariant under electromagnetic $U(1)$ gauge transformations (even in the presence of a boundary). The gauge field $a_\mu$ is dual to the electromagnetic current $J^\mu$, i.e.,
\beq
J^\mu= -\frac{\delta S_{QH}}{\delta A_\mu}=\frac{1}{2\pi}\epsilon^{\mu\nu\lambda}\partial_\nu a_\lambda.
\eeq

We consider a quantum Hall fluid with a flat edge along the cartesian coordinate $x$ (see Fig. \ref{fig1}).
To find the edge action, we first impose in the bulk a gauge fixing condition for the statistical gauge field
\begin{equation} \label{gaugefix}
a_t+c a_x=0.
\end{equation}
As we will demonstrate in Sec. \ref{Galsec}, this choice has the virtue of being invariant under Galilean boosts parallel to the boundary.  
In addition, the incompressibility condition $2\pi J^0=b=-\nu B$ that follows from the Gauss law can be automatically satisfied by writing
\begin{equation}\label{solai}
a_i=\partial_i\theta-\nu A_i.
\end{equation}
By substituting now Eqs. \eqref{gaugefix} and \eqref{solai} into the bulk action \eqref{QH} and integrating several times by parts one obtains a theory that lives only on the boundary with the action that can be most conveniently written in the following form\footnote{This is not how the action was originally written by Stone \cite{Stone1991}, but one can show that his result is equivalent to Eq. \eqref{StoneS} by using integration by parts.}
\begin{equation}\label{StoneS}
S_\theta=\frac{1}{4\pi}\int d^2 x\,  \left[ \frac{1}{\nu}D_+ \theta D_x\theta-\theta E_x\right], 
\end{equation}
where the covariant derivative
\begin{equation}
D_\mu\theta\equiv\partial_\mu\theta-\nu A_\mu
\end{equation}
was introduced and $D_+\theta \equiv D_t \theta +c D_x \theta$. Under the electromagnetic $U(1)$ gauge transformation $\delta_{\alpha} A_\mu=\partial_\mu\alpha$, the chiral boson is shifted
$\delta_\alpha \theta=\nu \alpha$
and thus the covariant derivative is invariant.
 Curiously, although there is no spontaneous symmetry breaking in the bulk of this system, the chiral boson couples to the external electromagnetic field as a Goldstone boson.  This is because it represents the phase of the vortex excitation, whose gap vanishes at the edge \cite{KaneFisher}. 

Due to the last term, the action \eqref{StoneS} is not $U(1)$ gauge invariant and thus has a gauge anomaly. In the presence of a boundary, however, the bulk Chern-Simons term
\begin{equation}\label{CS}
S_{CS}=\frac{\nu}{4\pi}\int d^3 x \,\epsilon^{\mu\nu\lambda}A_\mu \partial_\nu A_\lambda,
\end{equation}
that originates from Eq. \eqref{QH} by integrating out current and density fluctuations,
is also not gauge invariant with the variation given by the edge integral
\beq
\delta S_{CS}= \frac{\nu}{4\pi}\int d^2 x \,\alpha E_x.
\eeq
This contribution exactly cancels the gauge noninvariance of the edge theory \eqref{StoneS} and the total (bulk plus edge) theory is gauge invariant. This is a realization of the Callan-Harvey  anomaly inflow mechanism in quantum Hall physics \cite{Callan1985}.

\section{Galilean invariance} \label{Galsec}
In this paper we study clean quantum Hall fluids that are invariant under  rotations, translations and Galilean transformations. In an infinite flat system the bulk Chern-Simons action \eqref{CS} does not change under these spacetime transformations \cite{Dunne1999}. In the presence of a boundary, however, the rotation symmetry is lost and in addition the Chern-Simons action is not invariant anymore under Galilean boosts and translations in the direction perpendicular to the boundary. Nevertheless, Galilean boosts and translations along the boundary are still symmetries of the bulk CS action and thus need to be reflected by the edge action.

In this section we investigate how the edge theory \eqref{StoneS} behaves under Galilean boosts along a flat boundary. Let us recall that the transformations of the gauge fields $A_\mu$ and $a_\mu$ under a general infinitesimal spatial diffeomorphism $x^k\to x^k+\xi^k(t,\mathbf{x})$ and a $U(1)$ gauge transformation $\alpha$ are given by\footnote{As was found by Son and Wingate in \cite{Son2006}, this peculiar form of the transformation rule for the gauge potential $A_{\mu}$ can be most easily found by inspecting a free theory of nonrelativistic particles. There it was also demonstrated that microscopic interactions that respect general coordinate invariance can be introduced.} \cite{Son2006,Son2013}
\begin{equation}\label{NRdiffA}
\begin{split}
\delta A_t=& \dot{\alpha} -\xi^k\partial_k A_t-\dot{\xi}^k A_k,\\
\delta A_i =&\partial_i \alpha -\xi^k\partial_k A_i -\partial_i \xi^k A_k-m\dot{\xi}_i, \\
\delta a_t=& -\xi^k\partial_k a_t-\dot{\xi}^k a_k,\\
\delta a_i =& -\xi^k\partial_k a_i-\partial_i\xi^k a_k
\end{split}
\end{equation}
with $m$ denoting the mass of an elementary fermion (electron).\footnote{The convention used here differs from \cite{Son2006} by the sign of $m$. This convention is common in the recent works of Son and collaborators on quantum Hall physics \cite{Hoyos2012,Son2013,Geracie2014}.}  In fact, these transformation laws are valid provided the gyromagnetic factor $g_\psi$ and the spin $s_\psi$ of the elementary fermion satisfy the relation $g_\psi-2s_\psi=0$. This assumption can be easily relaxed, see \cite{Son2013,Geracie2014}.

A Galilean boost is a combination of the spatial diffeomorphism $\xi^k =\beta^k t$ and the $U(1)$ gauge transformation $\alpha= m\beta^k x_k$ with $\beta^k$ being the velocity parameter of the boost.
As a result, under Galilean boosts parallel to the boundary, the gauge fields $A_\mu$, $a_\mu$ and the chiral boson $\theta$ transform as
\begin{equation} \label{Gal}
\begin{split}
\delta_\beta A_t =&-\beta^x t \partial_x A_t-\beta^x A_x, \\
\delta_\beta A_i=&-\beta^x t\partial_x A_i,\\
\delta_\beta a_t =&-\beta^x t \partial_x a_t-\beta^x a_x, \\
\delta_\beta a_i=&-\beta^x t\partial_x a_i,\\ 
\delta_\beta \theta=&-\beta^x t\partial_x\theta+\nu m\beta^x x, 
\end{split}
\end{equation}
where we required that the chiral boson transforms as a Goldstone boson. It follows now from Eq. \eqref{Gal} that the magnetic and electric fields transform as
\begin{equation} \label{Gala}
\begin{split}
&\delta_\beta B =-\beta^x t \partial_x B,\\
& \delta_\beta E_x=-\beta^x t\partial_x E_x,\\
& \delta_\beta E_y=-\beta^x t \partial_x E_y-B \beta^x
\end{split}
\end{equation}
which implies the Galilean transformation law for the drift velocity \eqref{drift}
\beq \label{velxx}
\delta_\beta c=\beta^x.
\eeq
Using Eqs. \eqref{Gal}, \eqref{Gala}, \eqref{velxx}  we find that the edge action \eqref{StoneS} is not generically Galilean invariant. The variation of the action is given by
\begin{equation}
\delta_\beta S_\theta 
=- \frac{\nu }{2\pi}\int d^2 x\, m\beta^x A_+ 
\end{equation}
with $A_+\equiv A_t+c A_x$. Since the bulk Chern-Simons action \eqref{CS} is Galilean invariant, the total (bulk plus boundary) action is not!

The root of the problem can be identified by reexamining the derivation in Sec. \ref{CLL}: Although our starting point, the bulk action $S_{QH}$ \eqref{QH}, is invariant under Galilean boosts along the boundary, the identification \eqref{solai} is not consistent with Galilean invariance, since its left hand side and right hand side transform differently. Based on the transformation properties, we observe that Galilean invariance of the action can be recovered if we modify the $a_x$ component of the statistical gauge field to be
\begin{equation}
a_x=\partial_x \theta-\nu A_x-\nu m c.
\end{equation}
This identification is consistent with Galilean invariance and since $c$ is assumed to be constant at the edge, it is also consistent with the incompressibility constraint $b=-\nu B$. In the following sections we will put this idea on a firm ground using nonrelativistic general coordinate invariance introduced in \cite{Son2006}.

\section{Nonrelativistic general coordinate invariance}
In general, the form of the effective action is constrained by the (gauge) symmetries of the microscopic theory. It has been realized recently \cite{Hoyos2012,Son2013,Geracie2014} that in addition to the $U(1)$ gauge invariance, a large class of quantum Hall microscopic models are invariant under nonrelativistic general coordinate transformations (reparametrizations).  For this reason, it is important to incorporate this invariance into the effective action of quantum Hall fluids. In this section we review the basics of general coordinate invariance which will be used in the next section to construct the bulk and edge effective theories that respect this symmetry.

Consider a quantum Hall fluid living on an arbitrary two-dimensional manifold with a (generically time-dependent) spatial metric $g_{ij}$. The transformation of the spatial metric under an infinitesimal spatial diffeomorphism is given by
\begin{equation}
\delta g_{ij}=-\xi^k\partial_k g_{ij}-\partial_j \xi^k g_{ik}-\partial_i\xi^k g_{kj}.
\end{equation}
Since fermions in a quantum Hall fluid rotate and have a finite angular momentum, it is useful to introduce at every point of the manifold a pair of orthonormal spatial vectors (a vielbein) $e^a_i$ with $a=1,2$ that automatically satisfy
\beq \label{cond}
g_{ij}=e^a_i e^a_j, \quad \epsilon^{ab} e^a_i e^b_j= \varepsilon_{ij},
\eeq
where we introduced the Levi-Civita tensor $\varepsilon_{ij}\equiv \sqrt{g}\epsilon_{ij}$. 
Under spatial diffeomorphisms, the vielbein transforms as a one-form, i.e., 
\begin{equation}
\delta e_i^{a}=-\xi^k\partial_k e_i^{a}-\partial_i \xi^k e_k^{a}.
\end{equation}
The vielbein is not uniquely defined because it can be changed by a local spatial rotation acting on the frame indices $a,b$
\beq \label{trans}
e^a_i \to e^a_i - \lambda(t, \mathbf{x}) \epsilon^{ab} e^b_i
\eeq 
with the conditions \eqref{cond} preserved.
 We will denote the  group of such transformations as $SO(2)_V$\footnote{The subscript $V$ is added to distinguish from spatial rotations of the coordinates, that act on the $i,j$ indices. }. These transformations can be identified as rotations of the orbital spin.
This gauge freedom gives rise to the spin connection 
\begin{equation}
\begin{split}
\omega_t&= \frac{1}{2}\epsilon^{ab}e^{a k}\partial_t e_k^{b}, \\
\omega_i&= \frac{1}{2}\epsilon^{ab}e^{ak}\nabla_i e_k^{b}=\frac{1}{2} \left( \epsilon^{ab}e^{ak}\partial_i e_k^{b}-\varepsilon^{jk}\partial_j g_{ik} \right)
\end{split}
\end{equation}
which transforms as an Abelian gauge field $\omega_\mu\to \omega_\mu+ \partial_\mu \lambda$ under the local $SO(2)_V$ rotation \eqref{trans}. We will denote its electric and magnetic field as $E_{\omega\,i}$ and $B_\omega$, respectively.

Notably, the gauge potentials $A_\mu$ and $\omega_\mu$ do not transform as one-forms under spatial diffeomorphisms, see e.g. Eq. \eqref{NRdiffA}. This complicates the construction of general coordinate invariants. In the following we will use the modified fields
\begin{equation}
\begin{split}
&\tilde{A}_t=A_t-\frac{m}{2}g_{ij}v^i v^j, \ \ \tilde{A}_i=A_i+mg_{ij}v^j, \\
&\tilde{\omega}_t= \omega_t +\frac{1}{2}\varepsilon^{ij}\partial_i \left(g_{jk}v^k\right),\ \ \tilde{\omega}_i=\omega_i
\end{split}
\end{equation}
proposed in \cite{Hoyos2013, Son2013}. Here we introduced a velocity vector $v^i$ that transforms under spatial diffeomorphisms as
\begin{equation}
\delta v^i = -\xi^k\partial_k v^i+\partial_k \xi^i v^k+\dot{\xi}^i.
\end{equation}
As a result, $\tilde A_\mu$ and $\tilde \omega_\mu$ transform as one-forms for any choice of the velocity field. We use the modified gauge potentials to define modified electric and magnetic fields that transform covariantly under nonrelativistic diffeomorphisms 
$\tilde{E}_i,\tilde{B},\tilde{E}_{\omega\, i}$ and $\tilde{B}_\omega$.

Within this formalism one can automatically ensure Galilean invariance of a quantum Hall fluid provided $v^i$ can be written as a function of the spatial metric $g_{ij}$, the gauge potentials $A_\mu$ and their derivatives. For particles neutral under $SO(2)_V$ a natural choice for the velocity vector is given by \cite{Andreev2015}\footnote{This choice is physically natural because it gives the drift velocity of the Hall fluid in the presence of an external electric field. In a general case, the natural choice will involve the combination of the field strengths of the electromagnetic field and the spin connection under which particles are charged.}
\begin{equation} \label{tildev}
v^i=-\frac{\varepsilon^{ij}\tilde{E}_j}{\tilde{B}}.
\end{equation}
Note that the modified electromagnetic fields themselves depend on the velocity and its derivatives, so this has to be seen as an implicit non-linear differential equation for the velocity components. More explicitly, Eq. \eqref{tildev} for the velocity $v^i$ in terms of the electric and magnetic fields $E_i$ and $B$ is given by
\begin{equation}
(B+m\varepsilon^{kl}\partial_k v_l) v^i =-\varepsilon^{ij}(E_j+m\partial_t v_j+mv^k\partial_j v_k),
\end{equation}
which can be rewritten as the Euler equation for a charged fluid in the presence of electromagnetic fields
\begin{equation} \label{Euler}
-m(\partial_t+v^k\partial_k )v_i=E_i+Bv^k\varepsilon_{ki}.
\end{equation}
This equation will be solved order by order in derivatives in Sec. \ref{bulkac}.

\section{Bulk and edge theory} 
\subsection{Bulk action} \label{bulkac}
The effective action of a gapped quantum Hall fluid is a local functional of the external sources. It has an infinite number of terms that can be ordered according to a power-counting scheme. Here we utilize a derivative power-counting with a small parameter $\epsilon \ll 1$ 
\begin{equation}
A_t\sim 1,\ \ A_i\sim \frac{1}{\epsilon},\ \ e_i^{a}\sim g_{ij}\sim 1,\ \ \partial_i \sim \epsilon,\ \ \partial_t \sim \epsilon^2.
\end{equation}
This assignment makes the electromagnetic fields to be of order $B\sim 1$ and $E_i\sim \epsilon$. The Chern-Simons action \eqref{CS} is then the leading $O(1)$ order term.  Within this power-counting, the velocity field is expressed in terms of electromagnetic fields as follows: the Euler equation \eqref{Euler} is solved iteratively by expanding the velocity and solving order by order in $\epsilon$:
\begin{equation}
v^i=\epsilon v_0^i+\epsilon^3 v_1^i+\epsilon^5 v_2^i+\cdots,
\end{equation}
where the leading order solutions are
\begin{equation}
\begin{split}
v_0^i&=-\frac{\varepsilon^{ij}E_j}{B}, \\
v_1^i& =- \varepsilon^{ij}\frac{m}{B}(\partial_t+v_0^k\partial_k)v_{0\,j},\\
v_2^i&= \cdots.
\end{split}
\end{equation}

The next-to-leading $O(\epsilon^2)$ order corrections to the Chern-Simons bulk action were computed in \cite{Hoyos2012}. The bulk action  can be written as
\beq \label{sbulk}
S_{NLO}=\tilde S_{CS}+\tilde S_{WZ}+\tilde S_{\epsilon},
\eeq
where the general coordinate invariant form of the bulk Chern-Simons term is
\begin{equation}\label{SCScov}
\tilde S_{CS}=\frac{\nu}{4\pi}\int d^3x\, \epsilon^{\mu\nu\lambda} \tilde{A}_\mu \partial_\nu \tilde{A}_\lambda,
\end{equation}
the geometric general coordinate invariant Wen-Zee term \cite{wen1992} is
\begin{equation}\label{SWZ}
\tilde S_{WZ}=\frac{\nu s}{4\pi}\int d^3x\, \epsilon^{\mu\nu\lambda}\left( \tilde{\omega}_\mu \partial_\nu \tilde{A}_\lambda+\tilde{A}_\mu \partial_\nu \tilde{\omega}_\lambda \right)
\end{equation}
The coefficient is proportional to a number $s$ related to the orbital spin of particles in the ground state. It determines the ``shift'' $\mathcal{S}$, namely the mismatch between the number of fermions $N$ and number of elementary magnetic flux quanta $N_\phi$  for a quantum Hall fluid living on a closed manifold, i.e., $N_\phi=\nu^{-1}N-\mathcal{S}$ \cite{wen1992}. For a manifold of Euler characteristic $\chi_E$, the shift is $\mathcal{S}=s\chi_E$.

The Chern-Simons and Wen-Zee actions are topological. In general there are also non-topological terms that up to the order we are considering are
\begin{equation} \label{epsilon}
\tilde S_{\epsilon}=\int d^3x\,\sqrt{g}\left[ -\epsilon (\tilde{B})+K(\tilde{B})g^{ij}\partial_i \tilde{B}\partial_j \tilde{B}+h(\tilde{B}) R\right],
\end{equation}
where $\epsilon (\tilde{B})$, $K(\tilde{B})$ and $h(\tilde{B})$ are some functions of $\tilde B$ and $R$ denotes the Ricci scalar. Physically, $\epsilon(B)$ is the internal energy density of a quantum Hall fluid as a function of magnetic field.

Written in this form, the total action is manifestly general coordinate invariant. Although the action \eqref{sbulk} includes corrections of arbitrarily high order in $\epsilon$, it is complete only up to and including next-to-leading order.

\subsection{Edge action}
In the presence of a boundary we will consider a smaller subgroup of diffeomorphism transformations that preserve the shape of the boundary. In particular, for a boundary at $y=0$, this means that we will demand invariance of the edge action only under diffeomorphisms along the $x$ direction.

The form of the improved edge action can be derived from the \emph{topological action} of Wen and Zee \cite{wen1992} following the same steps as in Sec. \ref{CLL}. The general coordinate invariant form including the coupling to the spin connection is given by
\begin{equation} \label{tildeS}
\tilde S_{QH}=-\frac{1}{4\pi}\int d^3 x\, \epsilon^{\mu\nu\lambda}\left(\frac{1}{\nu}a_\mu\partial_\nu a_\lambda+2 a_\mu \partial_\nu \mathcal{\tilde A }_\lambda \right),
\end{equation}
where $\mathcal{\tilde A}_{\lambda}\equiv \tilde A_\lambda+s \tilde \omega_\lambda$.
In general, there are more \emph{non-topological terms} that depend on the dynamical gauge field $a_\mu$ (see for instance \cite{Son2013}) and give rise to Eq. \eqref{epsilon}. Since these terms do not affect the form of the edge theory, they have not been included here.

First we introduce a gauge fixing for the statistical field using the general coordinate invariant condition
\begin{equation}
a_t+ v^x a_x=0.
\end{equation}
This allows us to eliminate the temporal components of the gauge field.
On the other hand, its spatial components can be now written as
\begin{equation}
a_i=\partial_i\theta-\nu \mathcal{\tilde{A}}_i=\tilde{D}_i\theta.
\end{equation}
In this way, the Gauss law $b=-\nu \mathcal{\tilde B}=-\nu (\tilde{B}+ s \tilde{B}_\omega)$ is satisfied explicitly. As a result, one recovers the result that for a quantum Hall fluid on a sphere the shift is $\mathcal{S}=2s$.

 The action can be manipulated to be the sum of the bulk \emph{topological} terms \eqref{SCScov}, \eqref{SWZ} and the gravitational Chern-Simons term
\footnote{ The coefficient $\nu s^2$ is the classical value one gets from Eq. \eqref{tildeS}. Due to the framing anomaly, it will receive a quantum correction that matches the gravitational anomaly of the chiral theory at the boundary \cite{Gromov2014c,Gromov2015,Can2014}. This means that the quantum effective action of the chiral boson will have extra terms (in addition to Eq. \eqref{StoneScov}) that are not invariant under $SO(2)_V$ gauge transformations. These terms will compensate the variation of the bulk term, in such a way that the total action is invariant. This correction also determines the thermal Hall conductance \cite{Kane1997,Cappelli2001}. 
In this paper, however, we restrict the analysis only to the next-to-leading $O(\epsilon^2)$ order in the derivative expansion, and thus the gravitational Chern-Simons term and the associated extra $SO(2)_V$ anomalous terms in the edge action can be ignored.
}
 \begin{equation}\label{SGCS}
\tilde S_{GCS}=\frac{ \nu s^2}{4\pi}\int d^3x\, \epsilon^{\mu\nu\lambda}\tilde{\omega}_\mu \partial_\nu \tilde{\omega}_\lambda
\end{equation}
plus the boundary action for the chiral boson
\begin{equation}\label{StoneScov}
S_\theta=\frac{1}{4\pi}\int d^2 x\,  \left[ \frac{1}{\nu}\left(\tilde{D}_t \theta +v^x \tilde{D}_x \theta\right)\tilde{D}_x\theta-\theta \mathcal{\tilde{E}}_x \right],
\end{equation}
where $\mathcal{\tilde E}_x=\tilde E_x+s \tilde E_{\omega x}$. Here $x$ is a coordinate that parametrizes the boundary.
Eq. \eqref{StoneScov} is a key result of this paper that gives the generalization of the conventional edge action of Wen and Stone. 
The combination of this edge action with the bulk action \eqref{sbulk} is invariant under nonrelativistic diffeomorphisms, $U(1)$ and $SO(2)_V$ gauge transformations. It directly follows that the total action is Galilean invariant.

In this paper we assume the scaling $\theta\sim \epsilon^{-2}$. From Eq. \eqref{rhoeq}, this implies $\rho\sim \epsilon^{-1}$ which is a natural scaling for the one-dimensional fermion density at the edge. Within this power counting the Lagrangian of the edge theory \eqref{StoneScov} is $O(\epsilon^{-1})$. Note that in the absence of other considerations than symmetries, the action at this order can have a much more general form, the kinetic term can be an arbitrary function of $\tilde{D}_t\theta+v^x \tilde{D}_x\theta$ and the magnetic field $\tilde B$
\begin{equation}
S_\theta=\frac{1}{4\pi}\int d^2 x\,   \frac{1}{\nu}F\left(\tilde{D}_t \theta + v^x \tilde{D}_x \theta,\tilde{B}\right)\tilde{D}_x\theta-\theta \mathcal{\tilde{E}}_x. 
\end{equation}
The equation of motion is modified, but the solutions are still chiral with the same velocity. In the absence of sources $A_x=A_t=0$ and constant $B$ and $v^x$ they take the form
\begin{equation}
F''\partial_x\theta(\partial_t+v^x\partial_x)^2\theta+2F'(\partial_t+v^x\partial_x)\partial_x\theta=0.
\end{equation}

 \section{Electromagnetic response at the edge}
As an application, in this section we calculate linear response to longitudinal and transverse electromagnetic perturbations in a quantum Hall fluid with a flat boundary.

\subsection{Equation of motion and currents}
First, consider a chiral Luttinger liquid living on a generic boundary of a two-dimensional manifold. The boundary has an induced metric $h$ which can be calculated from the metric $g_{ij}$ of two-dimensional space as follows
\beq
h=\partial_x x^i \partial_x x^j g_{ij},
\eeq
where $x^i(x)$ is a parametrization of the boundary.

The equation of motion derived from the action \eqref{StoneScov} is
\beq
\partial_x \tilde D_+ \theta=0
\eeq
or written in momentum space
\beq \label{eom}
(\omega-c p_x) \theta=i \nu \mathcal{\tilde A}_+.
\eeq

The current is proportional to the variation of the effective action with respect to the external gauge field. In the presence of an anomaly, the current obtained in this way is not invariant under gauge transformations of the external field,\footnote{For non-Abelian symmetries the current does not transform covariantly.} but rather it transforms according to Wess-Zumino consistency conditions \cite{Wess1971}. For this reason it is known as the {\em consistent} current.  The current can be made {\em covariant} by adding additional terms, as first discussed by Bardeen and Zumino \cite{Bardeen1984}.\footnote{A discussion of consistent versus covariant currents for chiral gauge theories can also be found in \cite{Banerjee1986}.} These terms cannot be derived from a local action, but we can use the anomaly inflow argument of Callan and Harvey \cite{Callan1985} to determine them from the bulk action (see below).

The $U(1)$ \emph{consistent} current defined by
\beq \label{curnon}
J^\mu\equiv - \frac{1}{\sqrt{h}} \frac{\delta S_{\theta}}{\delta A_\mu}
\eeq
is given by
\beq
\begin{split}
J^t&=\frac {1} {\sqrt{h}} \Big(\frac {1} {2\pi}\tilde D_x \theta+\frac{\nu}{4\pi} \mathcal{\tilde A}_x \Big), \\
J^x&=\frac {1} {\sqrt{h}} \Big(\frac{c}{2\pi} \tilde D_x \theta -\frac{\nu}{4\pi} \mathcal{\tilde A}_t \Big).
\end{split}
\eeq
As explained above, this current is not gauge invariant, but we can make it such \cite{Hughes2013}: Under a variation with respect to the gauge potential $A_{\mu}$, the bulk action \eqref{sbulk} changes as\footnote{In order to extract the boundary current the variation is taken with vanishing normal derivatives $\partial_y \delta A_\mu=0$.  Otherwise the terms proportional to  $\partial_y \delta A_x\sim \delta B$ and $\partial_y \delta A_t\sim \delta E_y$ give the boundary magnetization and transverse polarization respectively, defined in the bulk as the variation of the grand canonical potential with respect to magnetic and electric fields. An example where the boundary magnetization is non-zero can be found in \cite{Hartnoll:2007ih} for free relativistic bosons in a magnetic field.}
\beq
\delta S_{NLO}=-\int d^3x \sqrt{g} J^{\mu}_{bulk} \delta A_\mu-\int d^2x \sqrt{h} J^{\mu}_{bound} \delta A_\mu.
\eeq

From the gauge invariance of the total (bulk plus boundary) action we find that the boundary current
\beq
\mathcal{\tilde J}^{\mu}=J^{\mu}+ J^{\mu}_{bound}
\eeq
must be gauge invariant. This is known as the \emph{covariant} current.
 The resulting covariant current is given by 
\beq \label{covcurr}
\begin{split}
\mathcal{\tilde J}^t&=\frac {1} {2\pi} \frac{1}{\sqrt{h}} \tilde D_x \theta+\mathcal{O}(\epsilon^2), \\
\mathcal {\tilde J}^x&=\frac{c}{2\pi} \frac{1}{\sqrt{h}} \tilde D_x \theta- \frac{1}{\sqrt{h}} \epsilon'(\tilde B)+\mathcal{O}(\epsilon^3),
\end{split}
\eeq
where $\epsilon'\equiv \partial_{\tilde B} \epsilon$. Here we discarded all terms that originate from $K(\tilde B)$ and $h(\tilde B)$ terms in the bulk action \eqref{epsilon} because they will introduce only higher order corrections in our power-counting. In addition, we assumed that the magnetic field is constant. This is sufficient for the calculation of electromagnetic linear response that is done in the following.


Using the equation of motion, one finds that the covariant current satisfies
\beq \label{imp}
\partial_\mu (\sqrt{h} \mathcal{ \tilde J}^\mu)
=-\frac{\nu}{2\pi} \mathcal{\tilde E}_x-\epsilon''(\tilde B) \partial_x \tilde B.
\eeq
The nonconservation of the current is a manifestation of the  \emph{covariant gauge anomaly}. 
\subsection{Longitudinal perturbation}
\begin{figure}[ht]
\begin{center}
\includegraphics[height=0.25\textwidth]{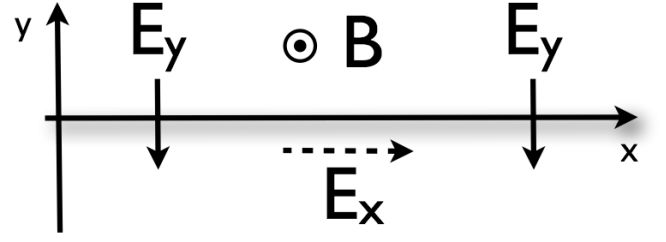}
\caption{Longitudinal perturbation $E_x$ (dashed line)}\label{fig2}
\end{center}
\end{figure}

We discuss now the case of a perturbation $\delta A_\mu(t,x)$ with $\mu=t,x$
in flat two-dimensional space with a flat boundary parametrized by a cartesian coordinate $x$. This perturbation leaves $B$ and $E_y$ unaffected, but gives rise to the nonvanishing longitudinal $E_x(t,x)$ (see Fig. \ref{fig2}). 

First, we calculate the electric conductivity and the conductance. By substituting the equation of motion \eqref{eom} into the expression for the covariant current \eqref{covcurr} we find in momentum space
\beq \label{deltaJ}
\begin{split}
&\delta \mathcal{\tilde J}^x(\omega, p_x)= \\
&\int_{p_y} \Big[\frac{\nu}{2 \pi} \frac{-i c}{\omega- c p_x} \mathcal{\tilde E}_x(\omega, \mathbf{p})- i m  \epsilon''(B) \epsilon^{kl} p_k v_l \Big],
\end{split}
\eeq
where $\int_{p_y}\equiv \int dp_{y}/(2\pi)$.

The velocity field in this case is given by
\beq \label{vel}
v^i=-\epsilon^{ij} \frac{\tilde E_j}{\tilde B} \approx \Big( c, \frac{E_x(t,x)}{B} \Big).
\eeq 

By substituting this into the general relation valid in flat space
\beq \label{gen}
\mathcal{\tilde E}_x=E_x+m \partial_t v_x+ \frac{m}{2}\partial_x \mathbf{v}^2-\frac{s}{2} \partial_x \epsilon^{ij} \partial_i v_j
\eeq
to linear order in external perturbation we find
\beq \label{tilde}
\mathcal{\tilde E}_x\approx \Big( 1 -\frac{\mathcal{S}}{4} \frac{\partial^2_x}{B} \Big) E_x,
\eeq
where the relation $2s=\mathcal{S}$ was used.

With the help of Eqs. \eqref{deltaJ}, \eqref{vel} and \eqref{tilde} we find now the electric conductivity $\sigma(\omega, p_x)$ in the edge theory
\beq
\begin{split}
&\delta \mathcal{\tilde J}^x(\omega, p_x)=\sigma(\omega, p_x) E_x(\omega, p_x), \\
&\sigma = \frac{\nu}{2 \pi} \Big( 1 +\frac{\mathcal{S}}{4} \frac{p^2_x}{B} \Big) \frac{-i c }{\omega- c p_x+i0^+} -im \epsilon''(B)\frac{p_x}{B}.
\end{split}
\eeq

In the limit $m\to 0$, the cyclotron frequency $\omega_c=B/m\to \infty$ and states from all higher Landau levels decouple. In this limit the conductivity simplifies to
\beq
\sigma(\omega, p_x)=\frac{\nu}{2 \pi} \Big( 1 +\frac{\mathcal{S}}{4} \frac{p^2_x}{B} \Big) \frac{-i c }{\omega- c p_x+i0^+} -i\frac{\nu}{2\pi}\frac{p_x}{B},
\eeq
where we used that $\epsilon(B)=\nu B^2/(4\pi m)$ for Laughlin's quantum Hall states in the limit $m\to 0$.

If one defines the conductance $G_H$ as a response of the current $\mathcal{\tilde J}^x$ to a constant $A_t$
\beq
\delta \mathcal{\tilde J}^x=G_H A_t,
\eeq
then
\beq
G_H= \frac{\nu}{2\pi}
\eeq
which agrees with the prediction of the chiral edge theory of \cite{Wen1992a,Stone1991}.

Finally, we note that the right-hand-side of Eq. \eqref{imp} has to be compensated by the bulk Hall current via the Callan-Harvey anomaly inflow. 
For a static perturbation  $\delta A_t(x)$
it was found in \cite{Hoyos2012} that
\beq
J^y=\frac{\nu}{2\pi} \Big(1-\frac{\mathcal{S}}{4} \frac{\partial^2_x }{B} \Big) E_x+ m \epsilon''(B)\frac{\partial^2_x E_x}{B}.
\eeq
This expression equals in magnitude and is opposite in sign to the right-hand-side of Eq. \eqref{imp}. The total $U(1)$ current is thus conserved in the bulk plus boundary system.

\subsection{Transverse perturbation}
\begin{figure}[ht]
\begin{center}
\includegraphics[height=0.25\textwidth]{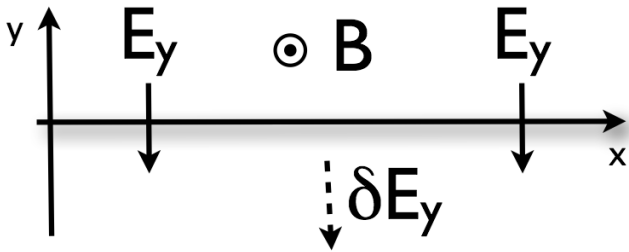}
\caption{Transverse perturbation $\delta E_y$ (dashed line)}\label{fig3}
\end{center}
\end{figure}

Consider now a perturbation $\delta A_\mu(t,y)$ with $\mu=t,y$
in flat two-dimensional space with a flat boundary along $x$. This perturbation leaves $B$ and $E_x$ unaffected, but gives rise to the nonvanishing transverse variation $\delta E_y(t,y)$ (see Fig. \ref{fig3}). In the following we will assume that $\delta E_y(t,y=0)=0$, i.e., the perturbation vanishes at the boundary. This is a simplest realization of a non-ideal edge.

For this perturbation the velocity field is given by
\beq \label{vela}
v^i=-\epsilon^{ij} \frac{\tilde E_j}{\tilde B} \approx \Big( c-\frac{\delta E_y(t,y)}{B}, 0  \Big).
\eeq 
Using Eqs. \eqref{gen} and \eqref{vela}, to linear order in the perturbation we find
\beq \label{tildeaa}
\mathcal{\tilde E}_x\approx -m  \frac{\partial_t \delta E_y(t,y)}{B}.
\eeq

Now the transverse conductivity $\sigma_H(\omega, \mathbf{p})$ will be calculated. In other words, we will determine the response of the edge covariant current $\delta \mathcal{\tilde J}^x$ to the transverse electric perturbation $\delta E_y$. 
Using Eqs. \eqref{deltaJ}, \eqref{vela} and \eqref{tildeaa} one gets
\beq 
\begin{split}
&\delta \mathcal{\tilde J}^x(\omega, p_x) = \int_{p_y} \sigma_H(\omega, \mathbf{p}) \delta E_y(\omega, p_y), \\
&\sigma_H(\omega, \mathbf{p})= \frac{\nu}{2 \pi} \frac{mc}{B} \frac{\omega }{\omega- c p_x+i0^+} -im \epsilon''(B)\frac{p_y}{B}.
\end{split}
\eeq
In the limit $m\to 0$ the transverse conductivity of a Laughlin state simplifies to
\beq
\sigma_H(\omega, p_y)=-i\frac{\nu}{2\pi}\frac{p_y}{B}.
\eeq

By using now the property of the perturbation
\beq
\delta E_y (t, y=0)= \int_{p_y} \delta E_y(\omega, p_y)=0,
\eeq
we find
\beq
\delta \mathcal{\tilde J}^x(\omega, p_x) =-im \epsilon''(B)\frac{\int_{p_y}p_y \delta E_y(\omega, p_y)}{B}
\eeq
which in the position space simplifies to
\beq
\delta \mathcal{\tilde J}^x(t, x) =-m \epsilon''(B)\frac{\partial_y \delta E_y(t,y)}{B} \Big|_{y=0}.
\eeq

\section{Outlook}
It has been demonstrated that \emph{Newton-Cartan geometry} is a powerful mathematical formalism to study nonrelativistic quantum Hall fluids,  superfluids and other nonrelativistic systems \cite{Christensen:2013lma, Son2013,Geracie2014,Gromov2015a,Banerjee2015,Brauner, Moroz2014a, Jensen2014}. In the future we plan to extend Newton-Cartan geometry to spacetime manifolds with boundaries, understand how it emerges from the bulk geometry and apply this formalism to the edges of quantum Hall fluids.

It is known that in addition to the electromagnetic gauge anomaly, the \emph{gravitational anomaly} appears at the edge of a quantum Hall fluid \cite{Kane1997, Cappelli2001}. While its consequences for the bulk of quantum Hall fluids have been studied extensively recently \cite{Can2014, Gromov2014c,Gromov2015}, it will be useful to investigate in more detail its edge implications by starting from the theory of a nonrelativistic general coordinate invariant chiral Luttinger liquid.


\section*{Acknowledgments:}
We acknowledge discussions with Victor Gurarie, Mike Hermele and Dam Thanh Son.
This research is supported by the U.S. Department of Energy (DOE), Office of Science, Basic Energy Sciences (BES) under Award \# DE-FG02-10ER46686 (S.M.).  This work is partially supported by the Spanish grant MINECO-13-FPA2012-35043-C02-02. C.H is supported by the Ramon y Cajal fellowship RYC-2012-10370. LR acknowledges support by the NSF grants DMR-1001240, and by the Simons Investigator award from the Simons Foundation.

\bibliography{library}

\end{document}